\documentclass[11pt,a4paper,oneside]{article}
\usepackage[T1]{fontenc}
\usepackage[utf8]{inputenc}
\usepackage{lmodern}
\usepackage[a4paper,margin=30mm]{geometry}
\usepackage{microtype}
\usepackage{amsmath,amssymb,amsthm}
\usepackage{graphicx}
\usepackage{booktabs,longtable,array}
\usepackage{calc}
\usepackage{xcolor}
\usepackage{bbm}
\usepackage[hidelinks]{hyperref}
\usepackage{orcidlink}
\usepackage{url}
\usepackage{titlesec}
\allowdisplaybreaks
\numberwithin{equation}{section}

\theoremstyle{definition}

\theoremstyle{remark}

\makeatletter
\newsavebox\pandoc@box
\newcommand*\pandocbounded[1]{\sbox\pandoc@box{#1}\ifdim\wd\pandoc@box>\linewidth\resizebox{\linewidth}{!}{\usebox\pandoc@box}\else\usebox\pandoc@box\fi}
\makeatother
\title{Classification of the Karlsson--Sz\"oll\H{o}si Intersection}
\author{André Ranchin\,\orcidlink{0009-0008-5789-9543}}
\date{}
\hypersetup{pdftitle={Classification of the Karlsson--Sz\"oll\H{o}si Intersection},pdfauthor={André Ranchin}}

\begin{document}
\maketitle
\begin{abstract}
We show that \ensuremath{K_6^{(2)}} $\cap$ \ensuremath{X_6^{(2)}} = \ensuremath{D_6^{(1)}} $\cup$ \ensuremath{C_6^{(1)}} up to standard Hadamard equivalence, where \ensuremath{C_6^{(1)}} is a non-affine one-parameter family. This demonstrates that the intersection of the two-parameter families of Karlsson and Szöllősi is strictly larger than the known affine Di\c{t}\u{a} curve \ensuremath{D_6^{(1)}}. Our new parametrisation of the Szöllősi family \ensuremath{X_6^{(2)}} in Karlsson coordinates and our constructive classification of the Karlsson--Szöllősi intersection shed light on the intersection of block-reducible forms and hypocycloid constraints. This may support the search for mutually unbiased bases in composite dimensions.
\end{abstract}

\tableofcontents

\section*{Introduction}\label{sec:introduction}

The study of complex Hadamard matrices is of intrinsic interest, has
applications in quantum information theory and is critical for finding mutually unbiased bases in composite dimensions \cite{McNultyWeigert2026, TadejZyczkowski2006}. The
complete classification of complex Hadamard matrices of order six
\cite{CardenesWuttigTindall2026} is an important step forward, but a more granular analysis of
non-affine Hadamard families is also likely to be useful in
understanding mutually unbiased configurations in dimension six\footnote{OpenAI released two papers dated September 2026 which claim to prove that N(6)=3. However, this work provided limited insight on the structure of mutually unbiased bases or Hadamard matrices.}.

In this paper, we study the intersection of the two-parameter families
of complex Hadamard matrices in dimension 6 discovered by Karlsson
\cite{Karlsson2009} and Szöllősi \cite{Szollosi2010}. It is known that they coincide along the
one-parameter affine curve arising from Di\c{t}\u{a}'s work \cite{Dita2004}, so that
\ensuremath{D_6^{(1)}} $\subseteq$ \ensuremath{K_6^{(2)}} $\cap$ \ensuremath{X_6^{(2)}}. In this paper, we demonstrate that this inclusion
is strict by introducing a non-affine one-parameter family in \ensuremath{K_6^{(2)}}
$\cap$ \ensuremath{X_6^{(2)}} which we call \ensuremath{C_6^{(1)}}.

We start with a few preliminaries in Section 1, then explicitly set out
a new parametrisation for the Szöllősi \ensuremath{X_6^{(2)}} family in terms of
Karlsson's three-parameter family \ensuremath{K_6^{(3)}} from \cite{Karlsson2011H2, Karlsson2011ThreeParameter}
in Section 2. We then define the \ensuremath{C_6^{(1)}} curve in Section 3,
which is not equivalent to \ensuremath{D_6^{(1)}}, and show that it is in the
intersection of \ensuremath{K_6^{(2)}} and \ensuremath{X_6^{(2)}}. We
introduce three technical lemmas in Section 4 which give a full classification of the intersection of \ensuremath{K_6^{(2)}} and \ensuremath{X_6^{(2)}}, with the proofs set out in Appendices A-C. The main theorem of this paper,
presented in Section 5, is that up to standard Hadamard equivalence
\ensuremath{D_6^{(1)}} $\cup$ \ensuremath{C_6^{(1)}} exhausts the full intersection of the two-parameter
Karlsson \ensuremath{K_6^{(2)}} and Szöllősi \ensuremath{X_6^{(2)}} families.

\section{Preliminaries}\label{sec:preliminaries}

We begin with the definitions needed to formulate the intersection classification. Readers familiar with the Karlsson and Szöllősi families
may prefer to skip directly to Section 2.

\textbf{Definition 1.1 (complex Hadamard matrix).} An
\(N \times N\)  matrix \(H = \left( h_{ij} \right)\ \) is a complex Hadamard matrix if:

\begin{equation}
 \mid h_{ij} \mid = 1\ \text{for~all~}i,j\ and\ HH^{\dagger} = N \mathbbm{1}_N 
\end{equation}

\textbf{Definition 1.2 (standard Hadamard equivalence).} Two complex
Hadamard matrices H and H' of the same order are \textbf{equivalent}
if:~

\begin{equation}
H' = D_r \ P_r \ H \ P_c \ D_c \ \ \ \
\end{equation}

where P$_r$ and P$_c$ are permutation matrices and D$_r$ and D$_c$ are diagonal unitary matrices. These operations permute and rephase rows and columns,
preserving the complex Hadamard property \cite{TadejZyczkowski2006}. Throughout this paper,
``equivalence'' means standard Hadamard equivalence.~

A complex Hadamard matrix is \textbf{dephased} if its first row and
first column consist entirely of 1s. Every standard-equivalence class
has a dephased representative.

Throughout this paper, a named Hadamard family  denotes a family of standard-equivalence classes. A displayed parametrised matrix typically
denotes a particular representative of the corresponding class.

\textbf{Definition 1.3 (affine family).} A Hadamard family is called \textbf{affine} if it admits a parametrised representative \(H(\theta_1,\ldots,\theta_r)\), with real parameters \(\theta_1,\ldots,\theta_r\), such that the argument of each matrix entry is an affine-linear function of the parameters: \begin{equation} H_{jk}(\theta_1,\ldots,\theta_r) = \exp\!\left( i\left( a_{jk,0} + a_{jk,1}\theta_1 + \cdots + a_{jk,r}\theta_r \right) \right) \end{equation} where the coefficients \(a_{jk,0},\ldots,a_{jk,r}\) are fixed real numbers and the arguments are understood modulo \(2\pi\). 

\textbf{Definition 1.4 (Karlsson's $H_2$-reducible \ensuremath{K_6}\textsuperscript{(3)}
family).} An order-six complex Hadamard matrix is $H_2$-reducible if it is
equivalent to a matrix admitting a 3$\times$3 block decomposition in which all
nine 2$\times$2 blocks are Hadamard. 
\newline Karlsson \cite{Karlsson2011H2} proved that an order-six complex Hadamard matrix is $H_2$-reducible if and only if it is equivalent to a matrix containing a 2 $\times$ 2 complex Hadamard
submatrix.
Karlsson \cite{Karlsson2011ThreeParameter} gave the complete closed-form parametrisation of the
$H_2$-reducible sector set out below. \newpage

Let:

\begin{equation}
F_{2} = \begin{pmatrix}
1 & 1 \\
1 & - 1
\end{pmatrix}\ \ \ \ \ \ \ \ \
\end{equation}

For phases \(z_{1},z_{2},z_{3},z_{4}\) on the complex unit circle
define:

\begin{equation}
\begin{matrix}
 & Z_{1} = \begin{pmatrix}
1 & 1 \\
z_{1} & - z_{1}
\end{pmatrix},\ Z_{2} = \begin{pmatrix}
1 & 1 \\
z_{2} & - z_{2}
\end{pmatrix}, 
\end{matrix} \\
\begin{matrix}
 & Z_{3} = \begin{pmatrix}
1 & z_{3} \\
1 & - z_{3}
\end{pmatrix},\ Z_{4} = \begin{pmatrix}
1 & z_{4} \\
1 & - z_{4}
\end{pmatrix}. 
\end{matrix}
\end{equation}

An $H_2$-reducible matrix can then be represented in the form:~

\begin{equation}
\begin{matrix}
 & K = \begin{pmatrix}
F_{2} & Z_{1} & Z_{2} \\
Z_{3} & \frac{1}{2}Z_{3}AZ_{1} & \frac{1}{2}Z_{3}BZ_{2} \\
Z_{4} & \frac{1}{2}Z_{4}BZ_{1} & \frac{1}{2}Z_{4}AZ_{2}
\end{pmatrix},\ A + B = - F_{2} 
\end{matrix}
\end{equation}

\ensuremath{K_6^{(3)}} is the family of standard-equivalence classes
admitting a representative of the form K in (1.6).

The core matrices \(A\ \) and \(B\) are determined by two real parameters
\(\theta,\phi\):

\begin{equation}
\begin{matrix}
 & A = F_{2}\left( - \frac{1}{2} \mathbbm{1}_2  + \frac{i\sqrt{3}}{2}\Lambda \right),\ B = F_{2}\left( - \frac{1}{2} \mathbbm{1}_2 - \frac{i\sqrt{3}}{2}\Lambda \right), 
\end{matrix}
\end{equation}where:

\begin{equation}
\begin{matrix}
 & \Lambda\left( \theta,\phi \right) = \begin{pmatrix}
\cos\theta & e^{i\phi}\sin\theta \\
e^{- i\phi}\sin\theta & - \cos\theta
\end{pmatrix} 
\end{matrix}
\end{equation}

\(\Lambda\ \)is Hermitian and \(\theta \in [0,2\pi)\) , \(\phi \in [0,2\pi)\). The four phases
\(z_{1},z_{2},z_{3},z_{4}\ \)are not independent as their squares are
related by the Möbius transformations arising from the unimodularity
conditions on the four core blocks (see \cite{Karlsson2011ThreeParameter}). Thus, after choosing
\(\left( \theta,\phi \right)\) and one independent boundary phase, the
remaining boundary phases are determined.

\textbf{Definition 1.5 (Karlsson's two-parameter family
\ensuremath{K_6^{(2)}}).} The \textbf{\ensuremath{K_6^{(2)}} family}
\cite{Karlsson2009} is the two-parameter subfamily of
\(\mathbf{H}_{\mathbf{2}}\)-reducible matrices obtained by imposing:

\begin{equation}
\begin{matrix}
 & Z_{1} = Z_{2},\ Z_{3} = Z_{4}. 
\end{matrix}
\end{equation}

\(K_{6}^{(2)}\)is the family of equivalence classes represented by
Karlsson's two-parameter matrices:

\begin{equation}
\begin{matrix}
 & K\left( z_{1},z_{3} \right) = \begin{pmatrix}
F_{2} & Z_{1} & Z_{1} \\
Z_{3} & \frac{1}{2}Z_{3}AZ_{1} & \frac{1}{2}Z_{3}BZ_{1} \\
Z_{3} & \frac{1}{2}Z_{3}BZ_{1} & \frac{1}{2}Z_{3}AZ_{1}
\end{pmatrix}
\end{matrix}
\end{equation}

By construction this means that \ensuremath{K_6^{(2)} \subseteq K_6^{(3)}}.

\textbf{Definition 1.6 (Szöllősi's two-parameter family
\ensuremath{X_6}\textsuperscript{(2)}).} We let:

\begin{equation}
\begin{matrix}
 & A = \begin{pmatrix}
a & b & c \\
c & a & b \\
b & c & a
\end{pmatrix},\ B = \begin{pmatrix}
d & e & f \\
f & d & e \\
e & f & d
\end{pmatrix} 
\end{matrix}
\end{equation}

be \(3 \times 3\ \)circulant matrices where \(a,b,c,d,e,f\), are complex
of modulus one, and form:

\begin{equation}
\begin{matrix}
 & H = \begin{pmatrix}
A & B \\
B^{\dagger} & - A^{\dagger}
\end{pmatrix}. 
\end{matrix}
\end{equation}

Since \(A\ \)and \(B\ \)are circulant, the Hadamard conditions reduce to \cite{Szollosi2010}:

\begin{equation}
\frac{a}{b} + \frac{b}{c} + \frac{c}{a} + \frac{d}{e} + \frac{e}{f} + \frac{f}{d} = 0\ \ \ \ \ \ \
\end{equation}

Using equivalence, set \(a = d = 1\). The remaining phases
\(b, \ c, \ e, \ f\ \)then satisfy

\begin{equation}
 \frac{1}{b} + \frac{b}{c} + c + \frac{1}{e} + \frac{e}{f} + f = 0. 
\end{equation}

After dephasing, one obtains:

\begin{equation}
\begin{matrix}
 & X\left( b,c,e,f \right) = \begin{pmatrix}
1 & 1 & 1 & 1 & 1 & 1 \\
1 & \frac{1}{bc} & \frac{b}{c^{2}} & \frac{f}{c} & \frac{1}{ce} & \frac{e}{cf} \\
1 & \frac{c}{b^{2}} & \frac{1}{bc} & \frac{e}{b} & \frac{f}{be} & \frac{1}{bf} \\
1 & \frac{1}{bf} & \frac{1}{ce} & - 1 & - \frac{1}{ce} & - \frac{1}{bf} \\
1 & \frac{e}{b} & \frac{e}{cf} & - \frac{e}{b} & - 1 & - \frac{e}{cf} \\
1 & \frac{f}{be} & \frac{f}{c} & - \frac{f}{c} & - \frac{f}{be} & - 1
\end{pmatrix}. 
\end{matrix}
\end{equation}

We define Szöllősi's two-parameter family \(X_{6}^{(2)}\) \cite{Szollosi2010} to be the
equivalence classes represented by the matrices in (1.15), where b,c,e,f
satisfy (1.14). Although four unit phases occur, the single complex
constraint generically leaves only two real degrees of freedom. Note that \(X_{6}^{(2)}\) is closed under transposition. Karlsson \cite{Karlsson2011ThreeParameter} showed that: \ensuremath{X_6^{(2)} \subseteq K_6^{(3)}}.

\textbf{Definition 1.7 (The Di\c{t}\u{a} curve \ensuremath{D_6^{(1)}}).} Let
\(z \in \mathbb{ C }\) satisfy \(\mid z \mid = 1\) and define:

\begin{equation}
D(z) = \begin{pmatrix}
1 & 1 & 1 & 1 & 1 & 1 \\
1 & - 1 & i & - i & - i & i \\
1 & i & - 1 & iz & - iz & - i \\
1 & - i & iz^{- 1} & - 1 & i & - iz^{- 1} \\
1 & - i & - iz^{- 1} & i & - 1 & iz^{- 1} \\
1 & i & - i & - iz & iz & - 1
\end{pmatrix}\ \ \
\end{equation}

The matrix \(D(z)\) is dephased and satisfies
\(D(z)D(z)^{\dagger} = 6 \mathbbm{1}_{6}\). The Di\c{t}\u{a} curve \ensuremath{D_6^{(1)}} is the set of
equivalence classes containing a representative D(z). It is an affine
one-parameter family \cite{Dita2004}. Writing \(z = e^{it}\), every entry has
argument \(a + nt\), where \(a\) is a fixed real constant
and\(\ n \in \left\{ - 1,0,1 \right\}\), modulo \(2\pi\). We write \( D_6:=D(1) \) for the order-six Diţă matrix \cite{Dita2004}.

\textbf{Proposition 1.1:} Karlsson placed the affine curve \ensuremath{D_6^{(1)}} in the
two-parameter Karlsson family \ensuremath{K_6^{(2)}} \cite{Karlsson2009} and Szöllősi placed \ensuremath{D_6^{(1)}} in
the two-parameter Szöllősi family \ensuremath{X_6^{(2)}} \cite{Szollosi2010}. So, up to equivalence,
we have:
\begin{equation}
\ensuremath{D_6^{(1)}} \subseteq \ensuremath{K_6^{(2)}} \cap \ensuremath{X_6^{(2)}}~~~
\end{equation}
In this paper, we introduce a non-affine one-parameter family which
we call \ensuremath{C_6^{(1)}}. We first show that \ensuremath{C_6^{(1)}} $\subseteq$ \ensuremath{K_6^{(2)}} $\cap$ \ensuremath{X_6^{(2)}} up to
equivalence,~then show that together, \ensuremath{D_6^{(1)}} and \ensuremath{C_6^{(1)}} exhaust the
intersection of the two-parameter Karlsson and Szöllősi families. Figure
1.1 depicts the relationships between the main subfamilies of the
\ensuremath{K_6^{(3)}} Karlsson sector. Note that this is a simplified
and illustrative picture. \newpage

\textbf{Figure 1.1: Depiction of the Karlsson sector, including the main subfamilies}

\includegraphics[width=0.78\textwidth]{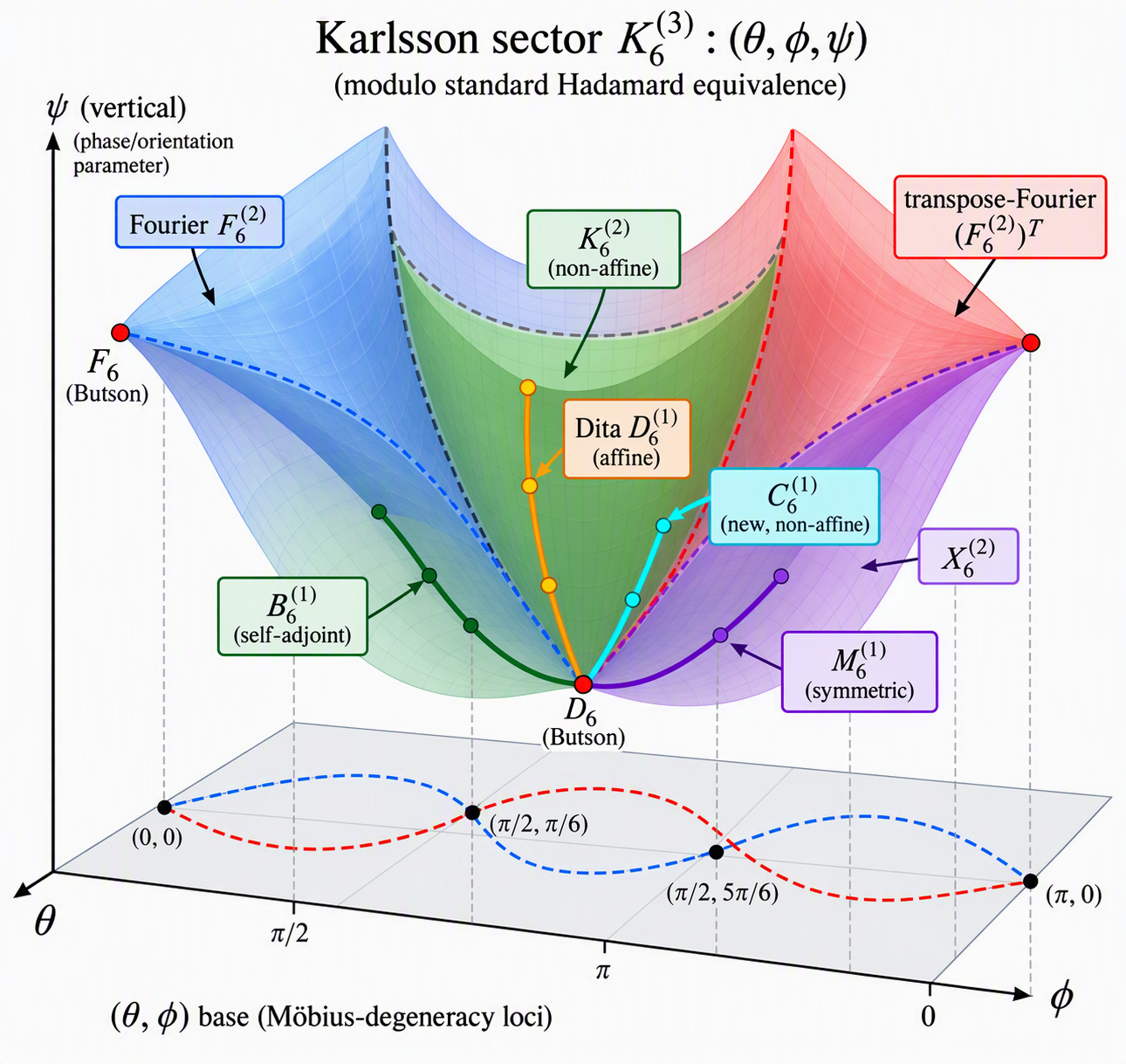}

\emph{Illustrative only. Note that distances, shapes, intersections and endpoint
locations are not drawn to scale and do not assert a global embedding of
the equivalence-class space.}

\section{\texorpdfstring{Karlsson parametrisation of \ensuremath{X_6^{(2)}}
}{Parametrisation of \ensuremath{X_6^{(2)}} within \ensuremath{K_6}(3) }}

Before studying the intersection of the two-parameter Karlsson and
Szöllősi families, we need to parametrise the Szöllősi \ensuremath{X_6^{(2)}} family in Karlsson block coordinates. This section explicitly sets out this
construction, which is split into three permutation charts.

Our initial objective is to put the X(b,c,e,f) matrices from (1.15) into
the Karlsson block form K from (1.6), reproduced below for convenience:

\begin{equation}
\begin{matrix}
 & K = \begin{pmatrix}
F_{2} & Z_{1} & Z_{2} \\
Z_{3} & \frac{1}{2}Z_{3}AZ_{1} & \frac{1}{2}Z_{3}BZ_{2} \\
Z_{4} & \frac{1}{2}Z_{4}BZ_{1} & \frac{1}{2}Z_{4}AZ_{2}
\end{pmatrix} 
\end{matrix}
\end{equation}

The three fixed diagonal entries \(X_{44} = X_{55} = X_{66} = - 1 \) in (1.15) give three canonical choices for pairing a fixed \(-1\) with \( X_{11} =1 \) to form the leading block \(F_{2} \). The permutations are chosen so that the four boundary blocks \( Z_{j}\) are in the format from (1.5). These are represented by the simultaneous row and column permutations:

\begin{equation}
{ K}_{j} = P_{j}X\left( b,c,e,f \right)P_{j}^{T} \end{equation}

where \( P_{j}\ \) is the permutation matrix for \( p_{j} \) and: 

\begin{equation}
\begin{aligned}
p_{1} & = \left( 1,4,2,6,3,5 \right) \\
p_{2} & = \left( 1,5,2,4,3,6 \right) \\
p_{3} & = \left( 1,6,2,5,3,4 \right) 
\end{aligned}
\end{equation}

where K\textsubscript{1}, K\textsubscript{2} and K\textsubscript{3} each
have Karlsson block form as per (2.1). Permuting the rows and columns of
\(X\left( b,c,e,f \right)\) and reading off the four \(2 \times 2\)
boundary blocks Z\textsubscript{1}, Z\textsubscript{2},
Z\textsubscript{3} and Z\textsubscript{4} from (1.5) then gives the table below:

\begin{equation}
\begin{matrix}
 & \begin{matrix}
\text{Pe}\text{rmutation} & z_{1} & z_{2} & z_{3} & z_{4} \\
p_{1} & (bf)^{- 1} & (ce)^{- 1} & f/c & e/b \\
p_{2} & e/b & e/(cf) & (ce)^{- 1} & f/(be) \\
p_{3} & f/(be) & f/c & e/(cf) & (bf)^{- 1}
\end{matrix} 
\end{matrix}
\end{equation}

In all three permutations, the four boundary phases satisfy:

\begin{equation}
\begin{matrix}
 & z_{1}z_{3} = z_{2}z_{4} = \frac{1}{bc}. 
\end{matrix}
\end{equation}

Working in the first permutation p\textsubscript{1}, we write:

\begin{equation}
x = z_{1} = \frac{1}{bf},\ y = z_{2} = \frac{1}{ce},\ z = z_{3} = \frac{f}{c}\ \ \ \
\end{equation}

Equation (2.5) then determines the fourth boundary phase:

\begin{equation}
\begin{matrix}
 & \left( z_{1},z_{2},z_{3},z_{4} \right) = \left( x,y,z,\frac{xz}{y} \right). 
\end{matrix}
\end{equation}

We can then recover the two-circulant parameters:

\begin{equation}
b = \frac{1}{xzc}, \ e = \frac{1}{yc}, \ f = zc
\end{equation}

and define: \(q := c^{3} \). Direct substitution into (1.14) gives:

\begin{equation}
\frac{1}{b} + \frac{b}{c} + c + \frac{1}{e} + \frac{e}{f} + f = \frac{q\text{\,}xyz(1 + xz + y + z) + x + y}{xyzc^{2}}\ \
\end{equation}

Because x, y, z and \(c^2\) are unit phases, the denominator is non-zero. So the two-circulant constraint is equivalent to the division-free equation:

\begin{equation}
\begin{matrix}
 & q\text{\,}xyz(1 + xz + y + z) + x + y = 0. 
\end{matrix}
\end{equation}

Conversely, let
x, y, z, q be complex numbers of modulus one that satisfy (2.10). Choose any unit phase c such that \(c^{3} = q\) and define \(b,e,f\) as in (2.8). Then \(b,c,e,f\) satisfy (1.14)
and hence determine a matrix \(X\left( b,c,e,f \right)\) in \(X_{6}^{(2)}\). The three
possible choices of \(c\) give exactly the same dephased matrix,
so the parametrisation is single valued at matrix level.

With respect to the permutation \(p_{1} = \left( 1,4,2,6,3,5 \right)\), the
resulting matrix is:

\begin{equation}
\begin{matrix}
 & K_{X}\left( x,y,z,q \right) = \begin{pmatrix}
1 & 1 & 1 & 1 & 1 & 1 \\
1 & - 1 & x & - x & y & - y \\
1 & z & xz & \frac{1}{yzq} & \frac{1}{xzq} & y \\
1 & - z & xyz^{2}q & - 1 & z & - xyz^{2}q \\
1 & \frac{xz}{y} & x^{2}z^{2}q & x & xz & xyz^{2}q \\
1 & - \frac{xz}{y} & \frac{xz}{y} & - \frac{1}{yzq} & \frac{1}{yzq} & - 1
\end{pmatrix} 
\end{matrix}
\end{equation}

Putting (2.11) explicitly into Karlsson block form then gives:

\begin{equation}
\begin{matrix}
 & K_{X} = \begin{pmatrix}
F_{2} & Z_{1}(x) & Z_{2}(y) \\
Z_{3}(z) & C_{11} & C_{12} \\
Z_{4}(xz/y) & C_{21} & C_{22}
\end{pmatrix} 
\end{matrix}
\end{equation}

Comparing (2.12) with (2.1) shows that the core blocks admit the Karlsson factorisation:

\begin{equation}
{C_{11} = \frac{1}{2}Z_{3}AZ_{1},\ C_{12} = \frac{1}{2}Z_{3}BZ_{2},
}\begin{matrix}
 & C_{21} = \frac{1}{2}Z_{4}BZ_{1},\ C_{22} = \frac{1}{2}Z_{4}AZ_{2}. 
\end{matrix}
\end{equation}

Thus (2.11) gives an explicit parametrisation of the standard
two-circulant family within \(K_{6}^{(3)}\). Finally, the specialisation
to the two-parameter Karlsson family \(K_{6}^{(2)}\) is obtained by
imposing, in a suitable block orientation:

\begin{equation}
Z_{1} = Z_{2},\ Z_{3} = Z_{4}\ \ \ \ \ \ \ 
\end{equation}

Note that, for the p\textsubscript{1} permutation, membership in \(K_{6}^{(2)}\) simply reduces
 to x=y.
\\

\textbf{Remark 2.1}: The \( p_1 \) permutation parametrises every standard dephased representative X(b,c,e,f) through (2.11) and (2.10). The permutations \( p_2 \) and \( p_3 \) give Karlsson block presentations of the same matrices. The recovery of the Karlsson  parametrisation from the \(p_1\) permutation only degenerates (i.e. fails to determine q) at:

\begin{equation}
\begin{matrix}
 & \left( x,y,z \right) = \left( \varepsilon, - \varepsilon, \varepsilon \right),\ \varepsilon \in \left\{ i, - i \right\}. 
\end{matrix}
\end{equation}

At every point of this indeterminate locus, at least one of the
p\textsubscript{2}, p\textsubscript{3} permutations remains regular. So the three permutations
together provide complete recovery of the standard parametrised \ensuremath{X_6^{(2)}}
family. At special parameter values, additional \(H_2 \)-reducible block decompositions may occur after more general row and column permutations; those conditional presentations are analysed in Section 4.

\section{The non-affine family \ensuremath{C_6^{(1)}}}\label{sec:new-family}

In this section, we formally define the \ensuremath{C_6^{(1)}} family, prove that
it is non-affine and show that it is distinct from the Di\c{t}\u{a} curve
\(D_{6}^{(1)}.\) We then prove that, up to Hadamard equivalence, \ensuremath{C_6^{(1)}}
is in the intersection of \ensuremath{K_6^{(2)}} and \ensuremath{X_6^{(2)}}.

\textbf{ Definition 3.1 (The
\(\mathbf{C}_{\mathbf{6}}^{\left( \mathbf{1} \right)}\ \)family).} Let $u, v \in \mathbb{C}$ with $\mid u \mid = \mid v \mid = 1$ such that:

\begin{equation}
\begin{matrix}
 & \left. \mid\left( \mathbf{1 + u} \right)\left( \mathbf{1 + v} \right) \right.\mid\mathbf{\  = 2} 
\end{matrix}
\end{equation}

Define:

\begin{equation}
\begin{matrix}
 & a = - \frac{(1 + u)(1 + v)}{2},\ d = - \frac{2uv}{(1 + u)(1 + v)}. 
\end{matrix}
\end{equation}

Since the left-hand side of (3.1) is non-zero, neither \(u\ \)nor
\(v\ \)is equal to \(- 1\), and the expressions in (3.2) are well
defined. Moreover,

\begin{equation}
\begin{matrix}
 & \mid a \mid = \mid d \mid = 1,\ ad = uv. 
\end{matrix}
\end{equation}

We define:

\begin{equation}
\begin{matrix}
 & \mathbf{C}\left( \mathbf{u},\mathbf{v} \right)\mathbf{=}\begin{pmatrix}
\mathbf{1} & \mathbf{1} & \mathbf{1} & \mathbf{1} & \mathbf{1} & \mathbf{1} \\
\mathbf{1} & \mathbf{uv} & \mathbf{a} & \mathbf{v} & \mathbf{u} & \mathbf{a} \\
\mathbf{1} & \mathbf{d} & \mathbf{uv} & \mathbf{v} & \mathbf{d} & \mathbf{u} \\
\mathbf{1} & \mathbf{u} & \mathbf{u} & \mathbf{- 1} & \mathbf{- u} & \mathbf{- u} \\
\mathbf{1} & \mathbf{v} & \mathbf{a} & \mathbf{- v} & \mathbf{- 1} & \mathbf{- a} \\
\mathbf{1} & \mathbf{d} & \mathbf{v} & \mathbf{- v} & \mathbf{- d} & \mathbf{- 1}
\end{pmatrix}\mathbf{.} 
\end{matrix}
\end{equation}

The \ensuremath{C_6^{(1)}} family is defined as the equivalence classes
represented by the matrices C\(\left( u,v \right)\ \)from (3.4), where u
and v satisfy (3.1). Proposition 3.3 will show that these are complex Hadamard matrices. Although two unit phases occur, (3.1) means that this is in fact a one-real-parameter family of Hadamard matrices. This is more transparent in
angular coordinates.

Write:

\begin{equation}
\begin{matrix}
 & u = e^{i\alpha},\ v = e^{i\beta},\ \alpha,\beta \in \left\lbrack - \pi,\pi \right\rbrack. 
\end{matrix}
\end{equation}

Then using \[ 1+e^{i\alpha} = 2e^{i\alpha/2}\cos\left(\frac{\alpha}{2}\right)\] we obtain:

\begin{equation}
\begin{matrix}
 & \left| (1 + u)(1 + v) \right| = 4\left| \cos\left( \frac{\alpha}{2} \right)\cos{(\frac{\beta}{2}}) \right| 
\end{matrix}
\end{equation}

Both half-angle cosines are non-negative for
\(\alpha,\beta \in \left\lbrack - \pi,\pi \right\rbrack\) so
(3.1) is equivalent to:

\begin{equation}
\begin{matrix}
 & \cos\frac{\alpha}{2}\cos\frac{\beta}{2} = \frac{1}{2}. 
\end{matrix}
\end{equation}

Using tangent half-angle substitutions:

\begin{equation}
\begin{matrix}
 & t = \tan\frac{\alpha}{2},\ s = \tan\frac{\beta}{2}, 
\end{matrix}
\end{equation}

we get:

\begin{equation}
\cos^{2}\frac{\alpha}{2} = \frac{1}{1 + t^{2}},\ \cos^{2}\frac{\beta}{2} = \frac{1}{1 + s^{2}}\
\end{equation}

which gives a real, algebraic curve equivalent to condition (3.1):

\begin{equation}
\begin{matrix}
 & F(t,s) = \ t^{2} + s^{2} + t^{2}s^{2} - 3 = 0 
\end{matrix}
\end{equation}

The curve is smooth as the gradient $\nabla$F=(2t(1+s\textsuperscript{2}),2s(1+t\textsuperscript{2})) can only vanish when t=s=0 which is not on the curve. So \ensuremath{C_6^{(1)}} is a
smooth one-dimensional family.\\

\textbf{Remark 3.1:}
The family $C_6^{(1)}$ is equivalent to the one-parameter subfamily $H^{(1,2)}$ identified in Case 3 of Proposition 5.3 of \cite{MatszangoszSzollosi2024}. Indeed, their constraint: \[ a^2-a+2ad+d^2-ad^2=0 \] is equivalent, under $u=-a$ and $v=-d^2/a$, to $(1+u)(1+v)=2d$, which yields the defining constraint (3.1). The corresponding matrices are equivalent via their two-circulant representatives. \newpage

\textbf{Proposition 3.1:}
\(\mathbf{C}_{\mathbf{6}}^{\left( \mathbf{1} \right)}\) \textbf{is a
non-affine family}

\textbf{Proof.}
Suppose, for contradiction, that \(C_{6}^{(1)}\) is affine in the sense of Definition~1.3. Since \(C_{6}^{(1)}\) is one-dimensional, a non-constant open arc then admits an affine representative \(H(t)\), defined on an open interval, with: 

\begin{equation} H_{jk}(t) = \exp\!\left(i(a_{jk}t+b_{jk})\right), \end{equation} 

where \(a_{jk}\) and \(b_{jk}\) are fixed real constants. Note that fixed row and column permutations preserve affineness by merely reordering the entrywise exponents, as does dephasing with respect to the first row and column, since it only subtracts affine-linear phase functions from entrywise exponents.

Let the  matrix C(u(t),v(t)) be obtained from H(t) by row and column permutations followed by dephasing. Since \(u(t)\) and \(v(t)\) occur as entries of \(C(u(t),v(t))\), we may choose continuous local arguments satisfying \begin{equation} u(t)=e^{i\alpha(t)}, \qquad v(t)=e^{i\beta(t)}, \end{equation} where \begin{equation} \alpha(t)=at+b, \qquad \beta(t)=ct+d \end{equation} for fixed real constants \(a,b,c,d\). 

The defining relation for \(C_{6}^{(1)}\) becomes:

\begin{equation} \cos\frac{\alpha(t)}{2} \cos\frac{\beta(t)}{2} = \frac{1}{2}.
\label{halfangle}
\end{equation} 

If \(a=0\), then \(\alpha\) is constant, and the preceding relation forces \(\beta\) to be constant. This contradicts the choice of a nonconstant arc. Hence \(a\neq0\), and eliminating \(t\) gives \begin{equation} \beta=m\alpha+n \end{equation} for fixed real constants \(m,n\). 

Substitution into \eqref{halfangle}, followed by the product-to-sum identity, gives: \begin{equation} \cos\frac{(1+m)\alpha+n}{2} + \cos\frac{(1-m)\alpha-n}{2} = 1. \end{equation} 

Differentiating twice with respect to \(\alpha\) and multiplying by \(-4\) gives: 
\begin{equation} (1+m)^2 \cos\frac{(1+m)\alpha+n}{2} + (1-m)^2 \cos\frac{(1-m)\alpha-n}{2} = 0. \end{equation} 

If \(m\notin\{-1,0,1\}\), the two cosine terms have distinct non-zero absolute frequencies and are therefore linearly independent on every open interval. Hence their linear combination cannot vanish identically, contradicting the preceding equation. If \(m=0\), then \(\beta\) is constant and the defining relation forces \(\alpha\) to be constant. If \(m=1\), then: \begin{equation} \cos\left(\alpha+\frac{n}{2}\right) + \cos\frac{n}{2} = 1, \end{equation} which cannot hold on an open interval. The case \(m=-1\) is identical, with the two cosine terms interchanged. Every possible value of m leads to a contradiction, which proves that \(C_{6}^{(1)}\) is non-affine. \qed

\newpage
\textbf{Proposition 3.2:} \textbf{ The family} \( \mathbf{ C_6^{(1)}}\) \textbf{is distinct from the Diţă family} \(\mathbf{ D_6^{(1)}}\)

\textbf{Proof.} The Diţă family \(D_6^{(1)}\) is affine \cite{Dita2004}, as it admits a dephased parametrisation in which the argument of every matrix entry is affine-linear in the parameter.

Suppose, for contradiction, that \( C_6^{(1)}=D_6^{(1)}\) as families of equivalence classes. The affine parametrisation of \(D_6^{(1)}\) would then also be an affine parametrisation of \(C_6^{(1)}\). This would contradict Proposition 3.1, which shows that \(C_6^{(1)}\) is non-affine in the sense of Definition 1.3. \qed 

\textbf{Remark 3.2:} The intersection of \( C_{6}^{(1)} \) and
\(D_{6}^{(1)}\) is non-empty as it contains the order-six Di\c{t}\u{a} matrix
\(D_6=D(1)\), which is equivalent to \(C(i,i)\). The exceptional Karlsson matrix $H_\times$ identified in \cite{CardenesWuttigTindall2026} is also equivalent to \(C(i,i)\) and \(D_6\).

More generally, in \(C_{6}^{(1)}\) each of the four parameter pairs below represents D\textsubscript{6}:

\begin{equation}
\begin{matrix}
 & \left( u,v \right) \in \left\{ \left( i,i \right),\left( i, - i \right),\left( - i,i \right),\left( - i, - i \right) \right\}
\end{matrix}
\end{equation}

\textbf{Proposition 3.3. Szöllősi and Karlsson membership of \ensuremath{C_6^{(1)}}}

Up to standard Hadamard equivalence, we have: \ensuremath{C_6^{(1)}} $\subseteq$
\(\ X_{6}^{(2)} \cap K_{6}^{(2)}.\)

\textbf{Proof.}

Let \(C\left( u,v \right)\) be a representative of \ensuremath{C_6^{(1)}} so that \(u,v\) are complex numbers of modulus one which satisfy:

\begin{equation}
\begin{matrix}
 & \left. \mid(1 + u) (1 + v)\mid = \right.2 
\end{matrix}
\end{equation}

Choose any complex number \(c\ \) satisfying:

\begin{equation}
 c^{3} = - \frac{2}{uv(1 + u)(1 + v)}
 \label{c_cube}
\end{equation}

and define:

\begin{equation}
\begin{matrix}
 & b = (uvc)^{- 1},\ e = (uc)^{- 1},\ f = vc. 
\end{matrix}
\end{equation}

The right-hand side of \eqref{c_cube} is unimodular, since:

\begin{equation}
 \left. \mid - \frac{2}{uv(1 + u)(1 + v)} \right.\mid = \frac{2}{\mid (1 + u)(1 + v) \mid} = 1. 
\end{equation}

We may therefore choose c to have modulus one, and then b, e, f also have modulus one.

We first prove membership in \(X_{6}^{(2)}\). Recall that the standard two-circulant representative \(X\left( b,c,e,f \right)\) is
defined for unit phases satisfying:

\begin{equation}
\begin{matrix}
 & \Phi\left( b,c,e,f \right) = \frac{1}{b} + \frac{b}{c} + c + \frac{1}{e} + \frac{e}{f} + f = 0. 
\end{matrix}
\end{equation}

Substituting the phases from (3.22) gives:

\begin{equation}
\begin{matrix}
 & \begin{aligned}
\Phi\left( b,c,e,f \right) & = uvc + \frac{1}{uvc^{2}} + c + uc + \frac{1}{uvc^{2}} + vc \\
 & = c(1 + u)(1 + v) + \frac{2}{uvc^{2}} \\
 & = \frac{uvc^{3}(1 + u)(1 + v) + 2}{uvc^{2}} \\
 & = 0
\end{aligned} 
\end{matrix}
\end{equation}

Thus \(b,c,e,f\) satisfy the defining two-circulant
constraint. Moreover, we have:

\begin{equation}
 \begin{array}{r}
(bc)^{- 1} = uv, \
\frac{f}{c} = \frac{e}{b} = v,\ (ce)^{- 1} = (bf)^{- 1} = u, \\
\frac{b}{c^{2}} = \frac{e}{cf} = - \frac{(1 + u)(1 + v)}{2} = a,\\ \frac{c}{b^{2}} = \frac{f}{be} = - \frac{2uv}{(1 + u)(1 + v)} = d.
\end{array} 
\end{equation}

Substitution into the standard representative
\(X\left( b,c,e,f \right)\ \) from (1.15) yields:

\begin{equation}
\begin{matrix}
 & X\left( b,c,e,f \right) = \begin{pmatrix}
1 & 1 & 1 & 1 & 1 & 1 \\
1 & uv & a & v & u & a \\
1 & d & uv & v & d & u \\
1 & u & u & - 1 & - u & - u \\
1 & v & a & - v & - 1 & - a \\
1 & d & v & - v & - d & - 1
\end{pmatrix} = C\left( u,v \right) 
\end{matrix}
\end{equation}

Therefore, the equivalence class represented by C(u,v) belongs to \(X_{6}^{(2)}\).

Next, we prove membership in \(K_{6}^{(2)}\). Let \(P_{1}\) be the
permutation matrix from Section 2 corresponding to the re-ordering $ p_{1} = \left( 1,4,2,6,3,5 \right)$.

As established in Section 2, the reblocked matrix:

\[K = P_{1}X\left( b,c,e,f \right)P_{1}^{T}
\]

has Karlsson block form, with boundary phases:
\begin{equation}
\begin{matrix}
 & z_{1} = (bf)^{- 1},\ z_{2} = (ce)^{- 1},\ z_{3} = \frac{f}{c},\ z_{4} = \frac{e}{b}. 
\end{matrix}
\end{equation}

The identities above therefore give:
\begin{equation}
  z_{1} = z_{2} = u,\ z_{3} = z_{4} = v. 
\end{equation}

Thus the reblocked matrix K has the repeated-boundary Karlsson form defining \(K_{6}^{(2)}\). Since $C(u,v)=X(b,c,e,f)$, the equivalence class represented by C(u,v) belongs to \(K_{6}^{(2)}\).

Therefore, \( C_6^{(1)}\subseteq X_6^{(2)}\cap K_6^{(2)} \) as claimed.\qed

\section{Classification of the \ensuremath{K_6^{(2)}} $\cap$ \ensuremath{X_6^{(2)}} intersection }\label{sec:classification}

The preceding sections described the standard two-circulant
representatives of \(X_{6}^{(2)}\) in Karlsson coordinates and
established that the family \(C_{6}^{(1)}\) is contained in \(K_{6}^{(2)} \cap X_{6}^{(2)}\).

It remains to prove the reverse inclusion:

\begin{equation} K_{6}^{(2)} \cap X_{6}^{(2)} \subseteq D_{6}^{(1)} \cup C_{6}^{(1)} \end{equation}

The three permutation charts from Section 2 provide natural Karlsson
presentations of a two-circulant representative \( X\left( b,c,e,f \right)\), but special parameter values may admit
additional block decompositions. To obtain a complete argument, we must
therefore consider every repeated-boundary presentation arising under
standard equivalence.

The inclusion proof is illustrated in Figure 4.1. Lemma 4.1 reduces all possible presentations to a finite system of frames, where every
frame is either \textbf{generic} or \textbf{conditional}. Lemma 4.2
identifies the generic frames with \(C_{6}^{(1)}\), while Lemma 4.3
shows that every non-empty conditional frame lies in
\(D_{6}^{(1)} \cup C_{6}^{(1)}\). The proofs are given in Appendices A--C.

\newpage
\textbf{Figure 4.1: proof structure of the intersection theorem}

\includegraphics[width=0.80\textwidth]{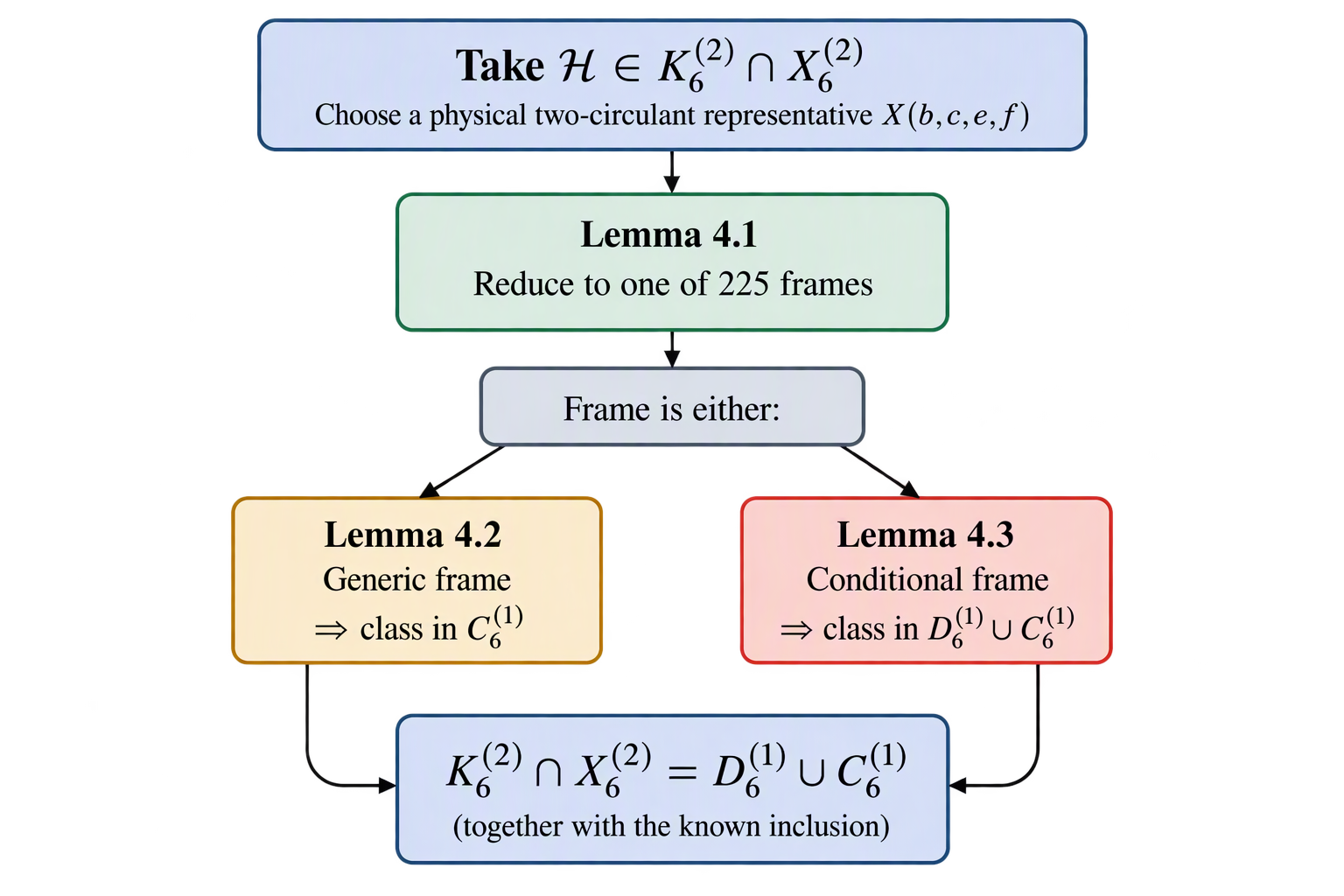}

\textbf{4.1 Finite-frame reduction}

\textbf{Definition 4.1 (physical representative).} A two-circulant representative
\(X = X\left( b,c,e,f \right)\) is called \textbf{physical} if \( \mid b \mid = \mid c \mid = \mid e \mid = \mid f \mid = 1\) and it
satisfies:
\begin{equation}
\Phi\left( b,c,e,f \right): = b^{- 1} + bc^{- 1} + c + e^{- 1} + ef^{- 1} + f = 0
\end{equation}

\textbf{Definition 4.2 (frames).} A \textbf{frame} consists of a partition of the
six row indices into three unordered pairs and, independently, a
partition of the six column indices into three unordered pairs. The two
partitions determine a \(3 \times 3\) array of nine \(2 \times 2\) blocks.

If a matrix equivalent to \(X\) has a repeated-boundary Karlsson
presentation, its block decomposition induces a frame of \(X\) by pulling
the row and column pairings back through the corresponding permutations.
We call this the \textbf{transported frame}.

Note that rephasings do not affect
either the pairings or the Hadamard property of the nine blocks. The
ordering, orientations and repeated-boundary equalities of the Karlsson
presentation are retained separately and are not part of the underlying
frame.

For a row pair \(\left\{ r,s \right\}\ \)and a column pair
\(\left\{ j,k \right\}\), define

\begin{equation}
\begin{matrix}
 & E_{rs;jk}(X): = X_{rj}X_{sk} + X_{rk}X_{sj}. 
\end{matrix}
\end{equation}

Since the entries of \(X\ \)are unimodular, the corresponding
\(2 \times 2\ \)block is Hadamard precisely when the block equations vanish:

\begin{equation}
E_{rs;jk}(X) = 0. 
\end{equation}

There are \(\frac{6!}{2^{3}\text{\,}3!} = 15\ \) pair partitions of six
indices and hence \(15^{2} = 225\ \)possible frames.

\textbf{Definition 4.3 (generic and conditional frames).} A frame is called \textbf{generic} if its nine block equations hold automatically for \(X(b,c,e,f)\), without imposing any additional condition on the parameters. A frame
that is not generic is called \textbf{conditional}.

A conditional frame's block equations impose additional restrictions on
the parameters \(b,c,e,f\). Such a frame may have no physical solutions,
or may occur only on a special locus. A conditional frame is
\textbf{empty} if it has no physical solution and \textbf{non-empty}
otherwise. 

\newpage
\textbf{Lemma 4.1 (finite-frame coverage)}

Let \(X = X\left( b,c,e,f \right)\) be physical and the equivalence class
represented by \(X\ \) belong to \(K_{6}^{(2)}\). Then every repeated-boundary Karlsson presentation equivalent to X induces one of the 225 frames of X, and all nine block equations associated with that frame vanish. \newline

Lemma 4.1 is a completeness reduction. The nine block equations identify
the underlying \(H_{2}\)-frame, but do not by themselves imply
membership in \(K_{6}^{(2)}\); the transported repeated-boundary
condition must also be retained.

The proof, given in Appendix A, uses the invariance of the vanishing of
\(E_{rs;jk}\) under row and column rephasing. Permutations transport both the row and column pairings and the repeated-boundary condition. No exceptional parameter values are lost in the process.

The table below sets out all 225  generic and conditional frames we need to consider.

\begin{longtable}{ @{} p{0.47\textwidth} >{\centering\arraybackslash}p{0.10\textwidth} >{\centering\arraybackslash}p{0.35\textwidth} @{} } \caption{Summary of the exact frame classification.} \label{tab:frame-classification}\\ \toprule \textbf{Frame class} & \textbf{Count} & \textbf{Outcome} \\ \midrule \endfirsthead \multicolumn{3}{c}{ \tablename\ \thetable{} -- continued from the previous page } \\ \toprule \textbf{Frame class} & \textbf{Count} & \textbf{Outcome} \\ \midrule \endhead \midrule \multicolumn{3}{r}{ Continued on the next page } \\ \endfoot \bottomrule \endlastfoot Generic & \(3\) & \(C_6^{(1)}\) for repeated-boundary \\ Empty conditional & \(168\) & No algebraic solution \\ Di\c{t}\u{a} conditional & \(27\) & \(D_6^{(1)}\) \\ Other positive-dimensional conditional & \(9\) & \(C_6^{(1)}\) \\ Zero-dimensional conditional & \(18\) & Points of \(C_6^{(1)}\) \\ \midrule \textbf{Total} & \(\mathbf{225}\) & \\ \end{longtable}

\textbf{4.2 Generic frames}

There are exactly three generic frames which
correspond to the permutations \(p_{1},p_{2},p_{3\ }\) introduced in Section 2. Lemma 4.2, proved in Appendix B, classifies the generic frames.

\textbf{Lemma 4.2 (generic frame identification)}

Let \(X=X(b,c,e,f)\) be a physical representative, and suppose that \(X\) admits an equivalent repeated-boundary Karlsson presentation supported on a generic frame. After normalising the choice of leading block and the ordering and orientation of the row and column pairs, the underlying oriented frame corresponds to one of the three permutations \(p_1,p_2,p_3\) introduced in Section~2. The repeated-boundary conditions are, respectively: 
\begin{equation} g_1:=bf-ce=0, \qquad g_2:=b-cf=0, \qquad g_3:=c-be=0. \end{equation} 

For each \(j\in\{1,2,3\}\), the physical locus \(g_j=0\) represents a member of \(C_6^{(1)}\).

So the equivalence class represented by X belongs to \( C_6^{(1)}\).

\textbf{4.3 Conditional frames}

We then need to treat conditional frames, whose block equations impose additional restrictions on \(b,c,e,f\). For each conditional frame, we consider the algebraic system of: 

\begin{enumerate} \item the two-circulant constraint, together with its conjugate relation; \item the nine block equations associated with the frame; and \item the non-vanishing conditions required for a physical representative. \end{enumerate}

Every physical solution of the underlying block-frame system lies in \( D_6^{(1)}\cup C_6^{(1)} \). \\

\textbf{Lemma 4.3 (conditional frame identification)}

Let \(X = X\left( b,c,e,f \right)\ \) be physical and suppose that a
repeated-boundary Karlsson presentation equivalent to \(X\ \)is
supported on a conditional frame. Then the equivalence class
represented by \(X\ \) belongs to:
\(D_{6}^{(1)} \cup C_{6}^{(1)}
\).

More precisely, the non-empty conditional frames comprise \(27\ \)Di\c{t}\u{a}
frames in \(D_{6}^{(1)}\), \(9\ \)positive-dimensional frames and
\(18\ \)zero-dimensional frames in \(C_{6}^{(1)}\). The proof, which exhaustively considers all the possible conditional frame types, is given in Appendix C.

\section{\texorpdfstring{The full Karlsson-Szöllősi intersection
}{The full intersection of the Karlsson and Szöllősi families }}\label{the-full-intersection-of-the-karlsson-and-szuxf6llux151si-families}

In this section, we prove our main theorem using the lemmas from the previous section.

\textbf{Theorem 5.1.} With the families understood as sets of
equivalence classes:

\[\mathbf{K}_{\mathbf{6}}^{\left( \mathbf{2} \right)} \ \mathbf{\cap} \ \mathbf{X}_{\mathbf{6}}^{\left( \mathbf{2} \right)} \ \mathbf{=} \ \mathbf{D}_{\mathbf{6}}^{\left( \mathbf{1} \right)} \ \mathbf{\cup} \ \mathbf{C}_{\mathbf{6}}^{\left( \mathbf{1} \right)}\]

\textbf{Proof.} The inclusion \(D_{6}^{(1)} \cup C_{6}^{(1)} \subseteq K_{6}^{(2)} \cap X_{6}^{(2)}  \) follows from the
previously established inclusion results for \(D_{6}^{(1)}\)and
\(C_{6}^{(1)}\) from Proposition 1.1 and Proposition 3.3 respectively.

For the reverse inclusion, let
\(\mathcal{H \in} K_{6}^{(2)} \cap X_{6}^{(2)}\) and choose a physical
standard two-circulant representative \(X\left( b,c,e,f \right)\) of
\(\mathcal{H}\). Lemma 4.1 places every equivalence class in
\(K_{6}^{(2)} \cap X_{6}^{(2)}\) within the finite 225 frame system. The frame is either generic or conditional.

Lemma 4.2 gives:
\[\text{generic~frame}\mathcal{\  \Longrightarrow \ H \in}C_{6}^{(1)},\]

Lemma 4.3 gives:
\[\text{conditional~frame}\mathcal{\  \Longrightarrow \ H \in}D_{6}^{(1)} \cup C_{6}^{(1)}.\]

Since these alternatives are exhaustive, we get:
\(K_{6}^{(2)} \cap X_{6}^{(2)} \subseteq D_{6}^{(1)} \cup C_{6}^{(1)}.  \)
 \qed

\section*{Conclusion}

The block-reducible construction of Karlsson and the circulant
construction of Szöllősi arise from markedly different structural
descriptions of complex Hadamard matrices. The main result of this paper
identifies precisely how these two constructions overlap, up to standard
Hadamard equivalence. In particular, the affine Di\c{t}\u{a}
curve does not exhaust the intersection: it is accompanied by the
non-affine family \(C_{6}^{(1)}\), which provides a second, structurally
distinct component.

The Fourier, Karlsson and
Szöllősi two-parameter families capture different symmetries and provide
complementary coordinate descriptions within the wider space of
order-six complex Hadamard matrices. Determining their intersections
helps distinguish genuine geometric relationships between these families
from apparent relationships arising from a particular choice of
parametrisation or representative.

Our initial motivation for this work was to better understand the
structure of mutually unbiased triples that overlap with non-affine
families of order 6. The intersection structure may help isolate areas
with additional symmetry or exceptional behaviour. Similar questions
could prove useful in higher composite dimensions, where several
inequivalent constructions coexist but the relationships between their
parameter spaces remain poorly understood. The present result
contributes to our understanding of the geometry of the order-six
Hadamard space and gives us some clues towards finding mutually unbiased
bases in composite dimensions.

\section*{Acknowledgements}

Thank you to Daniel McNulty and Karol {\.Z}yczkowski for their helpful comments.

OpenAI's GPT 5.6 Sol model was used in the development of this work. The model assisted with testing proof strategies, algebraic calculations, symbolic-code development, proof auditing, generating figures and editorial revision.

The author specified the definitions and target statements, supplied and checked the relevant literature, selected the proof strategies, reviewed the resulting derivations and computations, and decided what material to include. The author takes full responsibility for the mathematical content and the presentation of this work.

\bibliographystyle{unsrt}
\bibliography{KXreferences}

\clearpage 
\appendix 

\phantomsection \addcontentsline{toc}{section}{Appendices} 

\addtocontents{toc}{\protect\setcounter{tocdepth}{-1}}

\titleformat{\section} {\normalfont\Large\bfseries} {Appendix \thesection :} {1em} {}

\section{Proof of Lemma 4.1}\label{app:lemma41}

\textbf{Lemma 4.1 (finite-frame coverage)}

Let \(X = X\left( b,c,e,f \right)\) be physical as per Definition 4.1. If
the equivalence class represented by \(X\ \) belongs to \(K_{6}^{(2)}\),
then a repeated-boundary Karlsson presentation equivalent to
\(X\) transports to a presentation over one of the \(225\ \)frames of
\(X\). In the underlying frame, all nine block equations hold:

\begin{equation}
\begin{matrix}
 & E_{rs;jk}(X): = X_{rj}X_{sk} + X_{rk}X_{sj} = 0 
\end{matrix}
\end{equation}

An oriented framed presentation consists of ordered row pairs, ordered
column pairs, a choice of leading block, and orientations within each
pair.

The transported presentation retains the distinguished block pairs,
their ordering and orientations, and the repeated-boundary equalities.
This reduction requires no division by a parameter-dependent quantity.

\textbf{Proof of Lemma 4.1}

Let \(X = X\left( b,c,e,f \right)\) be physical, and suppose that its
equivalence class belongs to \(K_{6}^{(2)}\). By Definitions 1.2 and
1.5, there exist permutation matrices \(P_{r},P_{c}\) and diagonal
unitary matrices \(D_{r},D_{c}\ \) such that:

\begin{equation}
\begin{matrix}
 & K = D_{r}P_{r}XP_{c}D_{c} 
\end{matrix}
\end{equation}

is a repeated-boundary Karlsson representative.

The \(3 \times 3\) block decomposition of \(K\)partitions its row indices
into three unordered pairs and, independently, its column indices into
three unordered pairs. Transporting these partitions back through the
permutations in (A.2) gives a frame \(F\) of \(X\).

For a row pair $\{r,s\}$ and a column pair $\{j,k\}$, consider the 2 $\times$ 2 matrix block: 

\begin{equation} B= \begin{pmatrix} X_{rj} & X_{rk}\\ X_{sj} & X_{sk} \end{pmatrix}. \end{equation}

Since its entries are unimodular, $B$ is Hadamard if and only if the two rows are orthogonal: 
\begin{equation}
X_{rj}\overline{X_{sj}}+X_{rk}\overline{X_{sk}}=0. \end{equation} 

Multiplying by the non-zero quantity $X_{sj}X_{sk}$ and using $\overline{X_{ab}}=X_{ab}^{-1}$ shows that the Hadamard condition is equivalent to: 

\begin{equation} 
E_{rs;jk} = X_{rj}X_{sk}+X_{rk}X_{sj}=0. 
\end{equation} 

It remains to verify the nine block equations in \(F\). Row and column
permutations merely relabel the corresponding index pairs. Moreover, if:

\begin{equation}
Y = D_{\rho}XD_{\gamma},\ D_{\rho} = \operatorname{diag}\left( \rho_{1},\ldots,\rho_{6} \right),\ D_{\gamma} = \operatorname{diag}\left( \gamma_{1},\ldots,\gamma_{6} \right)
\end{equation}

then

\begin{equation}
\begin{matrix}
 & E_{rs;jk}(Y) = \rho_{r}\rho_{s}\gamma_{j}\gamma_{k}\text{\,}E_{rs;jk}(X). 
\end{matrix}
\end{equation}

The prefactor in (A.7) is non-zero, so the vanishing of \(E_{rs;jk}\) is
preserved under row and column rephasing. Since all nine
\(2 \times 2\ \)blocks of \(K\) are Hadamard, all nine equations:

\begin{equation}
E_{rs;jk}(X) = 0
\end{equation}
hold in the transported frame \(F\).

There are

\[\frac{6!}{2^{3}\text{\,}3!} = 15
\]

partitions of six indices into three unordered pairs. This is because 6! orders the six elements into three pairs; $2^3$ removes the orientation within each of the three pairs and 3! removes the ordering of the three pairs.

Since the row and
column partitions are independent, there are \(15^{2} = 225\ \)possible
frames. Hence \(F\) must be one of the 225 frames considered in Section 4.

Finally, the transported presentation retains the ordering and orientation of the distinguished block pairs, together with the equivalence data under which \(Z_1=Z_2\) and \(Z_3=Z_4\). These equalities hold in the resulting Karlsson representative, rather than necessarily between the corresponding subblocks of \(X\). Since no parameter-dependent division is used, the reduction remains valid when an individual Karlsson parametrisation is indeterminate. So every physical Szöllősi representative \(X\) whose equivalence class belongs to \(K_6^{(2)}\) is covered by at least one of the \(225\) frame systems. \qed

\section{Proof of Lemma 4.2}\label{app:lemma42}

\textbf{Lemma 4.2 (generic frame identification)}

Let \(X=X(b,c,e,f)\) be a physical representative, and suppose that \(X\) admits an equivalent repeated-boundary Karlsson presentation supported on a generic frame. After normalising the choice of leading block and the ordering and orientation of the row and column pairs, the underlying oriented frame corresponds to one of the three permutations \(p_1,p_2,p_3\) introduced in Section~2. The corresponding repeated-boundary conditions are, respectively: 

\begin{equation} g_1:=bf-ce=0, \qquad g_2:=b-cf=0, \qquad g_3:=c-be=0. \end{equation} 

For each \(j\in\{1,2,3\}\), the physical locus \(g_j=0\) represents a member of \(C_6^{(1)}\).

So the equivalence class represented by X belongs to \( C_6^{(1)}\).

\textbf{Proof of Lemma 4.2.} A frame is \emph{generic} if its nine block equations hold automatically for \(X(b,c,e,f)\), without imposing any additional condition on the parameters. Equivalently, after substituting the entries of \(X(b,c,e,f)\), the left-hand sides of all nine block equations (A.8) simplify algebraically to zero. Direct evaluation of all \(225\) frames shows that, up to the choice of leading block and the ordering and orientation of the row and column pairs, the generic frames are precisely those induced by the permutations \(p_1,p_2,p_3\) in (2.3). 

The corresponding boundary phases from (2.4) are: \begin{equation}\label{eq:generic-boundary-phases} \begin{array}{c|cccc} & z_1 & z_2 & z_3 & z_4\\ \hline p_1 & (bf)^{-1} & (ce)^{-1} & f/c & e/b\\ p_2 & e/b & e/(cf) & (ce)^{-1} & f/(be)\\ p_3 & f/(be) & f/c & e/(cf) & (bf)^{-1}. \end{array} \end{equation} 

A Karlsson presentation belongs to \(K_6^{(2)}\) precisely when \( z_1=z_2, \ z_3=z_4 \). Since \(b,c,e,f\) are non-zero, these two equalities reduce to the three charts with the conditions from (B.1).

In each chart, set \begin{equation} u:=z_1=z_2, \qquad v:=z_3=z_4. \end{equation}  

For the \(p_1\)-chart, this gives: \begin{equation} u=(bf)^{-1}=(ce)^{-1}, \qquad v=\frac{f}{c}=\frac{e}{b}. \end{equation} 

This is exactly the parametrisation from Proposition~3.3. There, the physical two-circulant constraint was shown to imply \begin{equation} (1+u)(1+v)=-\frac{2}{uvc^3}, \end{equation} and the corresponding standard two-circulant representative was identified with \(C(u,v)\).

For the \(p_2\)-chart, the condition \(g_2=0\) gives \(b=cf\), and we set: \begin{equation}\label{eq:p2-boundary-phases} u=\frac{e}{b}=\frac{e}{cf}, \qquad v=(ce)^{-1}=\frac{f}{be}. \end{equation} Solving these relations in terms of \(u,v,c\) gives \begin{equation}\label{eq:p2-substitution} b=\frac{1}{uvc}, \qquad e=\frac{1}{vc}, \qquad f=\frac{1}{uvc^2}. \end{equation} 

Substitution into the physical two-circulant constraint \begin{equation}\label{eq:physical-constraint-appB} \Phi(b,c,e,f) = \frac{1}{b}+\frac{b}{c}+c+\frac{1}{e}+\frac{e}{f}+f = 0 \end{equation} gives \begin{equation}\label{eq:p2-constraint-substitution} \Phi(b,c,e,f) = \frac{uvc^3(1+u)(1+v)+2}{uvc^2}. \end{equation} Hence \begin{equation}\label{eq:p2-C-relation} (1+u)(1+v)=-\frac{2}{uvc^3}. \end{equation} 

Since \(u,v,c\) are unit phases, it follows that \begin{equation}\label{eq:p2-C-modulus} \left|(1+u)(1+v)\right|=2. \end{equation}

Let \(P_{\sigma_2}\) be the permutation matrix corresponding to \begin{equation} \sigma_2=(1,2,3,5,6,4). \end{equation} Using \eqref{eq:p2-substitution} and \eqref{eq:p2-C-relation}, direct entrywise simplification for the \(p_2\)-chart gives: \begin{equation}\label{eq:p2-C-equivalence} P_{\sigma_2} X\left( \frac{1}{uvc}, c, \frac{1}{vc}, \frac{1}{uvc^2} \right) P_{\sigma_2}^{T} = C(u,v). \end{equation} Thus every physical solution of \(g_2=0\) represents a member of \(C_6^{(1)}\).

For the \(p_3\)-chart, the condition \(g_3=0\) gives \(c=be\), while \begin{equation}\label{eq:p3-boundary-phases} u=\frac{f}{be}=\frac{f}{c}, \qquad v=\frac{e}{cf}=(bf)^{-1}. \end{equation} Solving these relations in terms of \(u,v,b\) gives \begin{equation}\label{eq:p3-substitution} c=\frac{1}{ubv}, \qquad e=\frac{1}{ub^2v}, \qquad f=\frac{1}{bv}. \end{equation} 

Substitution into \eqref{eq:physical-constraint-appB} gives \begin{equation}\label{eq:p3-constraint-substitution} \Phi(b,c,e,f) = \frac{(1+u)(1+v)+2b^3u^2v^2}{buv}. \end{equation} Hence \begin{equation}\label{eq:p3-C-relation} (1+u)(1+v)=-2b^3u^2v^2. \end{equation} 

Since \(b,u,v\) are unit phases, it follows that \begin{equation}\label{eq:p3-C-modulus} \left|(1+u)(1+v)\right|=2. \end{equation}

Let \(P_{\sigma_3}\) be the permutation matrix corresponding to \begin{equation} \sigma_3=(1,2,3,6,4,5). \end{equation} Using \eqref{eq:p3-substitution} and \eqref{eq:p3-C-relation}, direct entrywise simplification for the \(p_3\) chart gives: \begin{equation}\label{eq:p3-C-equivalence} P_{\sigma_3} X\left( b, \frac{1}{ubv}, \frac{1}{ub^2v}, \frac{1}{bv} \right) P_{\sigma_3}^{T} = C(u,v). \end{equation} Thus every physical solution of \(g_3=0\) represents a member of \(C_6^{(1)}\).

The \(p_1\)-chart was identified with \(C(u,v)\) in Proposition~3.3, while \eqref{eq:p2-C-equivalence} and \eqref{eq:p3-C-equivalence} establish the corresponding identifications for the \(p_2\)- and \(p_3\)-charts. Consequently, every physical solution of \(g_1=0\), \(g_2=0\), or \(g_3=0\) represents a member of \(C_6^{(1)}\). Hence every physical matrix admitting an equivalent repeated-boundary Karlsson presentation supported on a generic frame represents a member of \(C_6^{(1)}\). \qed

\section{Proof of Lemma 4.3}\label{app:lemma43}

This appendix gives the exact finite classification underlying Table 1 and proves Lemma 4.3.  The proof is computer-assisted but exact: the enumerative steps are certified by reduced Gr\"obner bases and normal forms over a rational polynomial ring, while family membership is certified by explicit standard-equivalence identities.

\textbf{Lemma 4.3 (conditional frame identification)}

Let \(X = X\left( b,c,e,f \right)\ \) be physical and suppose that a
repeated-boundary Karlsson presentation equivalent to \(X\ \)is
supported on a conditional frame. Then the equivalence class
represented by \(X \) belongs to:
\(D_{6}^{(1)} \cup C_{6}^{(1)}
\).

\subsection{Frames and repeated-boundary presentations}
\label{appC:frames}

Let $X=X(b,c,e,f)$ be a physical representative as per Definition 4.1 so that:
\begin{equation}
 \Phi(b,c,e,f)
 :=b^{-1}+bc^{-1}+c+e^{-1}+ef^{-1}+f
 =0
 \label{eq:C-Phi}
\end{equation}

Complex conjugation sends~\eqref{eq:C-Phi} to
\begin{equation}
 \Phi^{\ast}(b,c,e,f)
 :=b+cb^{-1}+c^{-1}+e+fe^{-1}+f^{-1}
 =0
 \label{eq:C-Phistar}
\end{equation}

The second equation is automatic for physical parameters but is included explicitly in the algebraic system below.

A \emph{frame} specifies a perfect matching of the six rows and a perfect matching of the six columns. It therefore records three unordered row pairs and three unordered column pairs. The ordering and orientation are required to transport the repeated-boundary conditions: \begin{equation} Z_1=Z_2, \qquad Z_3=Z_4 \end{equation} which form part of the Karlsson presentation, rather than of the frame itself.

We order the fifteen perfect matchings as follows: 

\begin{equation} \begin{aligned} M_0&=(12)(34)(56), & M_1&=(12)(35)(46), & M_2&=(12)(36)(45),\\ M_3&=(13)(24)(56), & M_4&=(13)(25)(46), & M_5&=(13)(26)(45),\\ M_6&=(14)(23)(56), & M_7&=(14)(25)(36), & M_8&=(14)(26)(35),\\ M_9&=(15)(23)(46), & M_{10}&=(15)(24)(36), & M_{11}&=(15)(26)(34),\\ M_{12}&=(16)(23)(45), & M_{13}&=(16)(24)(35), & M_{14}&=(16)(25)(34). \end{aligned} \label{eq:C-matching-order} \end{equation} 

For \(0\leq r,s\leq14\), the frame 

\begin{equation} F_{rs}:=(M_r,M_s) \end{equation}

uses \(M_r\) for the rows and \(M_s\) for the columns. The fifteen choices in each direction give exactly \(15^2=225\) frames. In the following, we abbreviate \(F_{rs}\) to \((r,s)\). 

\subsection{The algebraic frame ideals} \label{appC:ideals} We associate with each frame an exact polynomial system consisting of the two-circulant equation, its conjugate, and the nine block equations from (A.1) determined by that frame. The entries of \(X\), and hence the expressions \(\Phi\), \(\Phi^{\ast}\) and \(E_{rs;jk}\), are Laurent polynomials in \(b,c,e,f\); that is, they may contain negative powers of these variables. For any such expression \(h\), let \(\widetilde h\) denote the numerator obtained after \(h\) has been written over a common denominator. 

Provided that \(b,c,e,f\) are non-zero, the equations: \begin{equation} h=0 \qquad\text{and}\qquad \widetilde h=0 \label{eq:C-cleared-equation} \end{equation} are equivalent. 

To retain the non-vanishing condition after clearing denominators, we introduce an auxiliary variable \(\tau\) and impose: \begin{equation} \tau bcef-1=0. \label{eq:C-nonzero-condition} \end{equation} This equation has a solution for \(\tau\) if and only if \(bcef\neq0\), in which case \begin{equation} \tau=(bcef)^{-1}. \label{eq:C-tau-definition} \end{equation} Since physical parameters lie on the unit circle, every physical solution has \(b,c,e,f\neq0\) and therefore determines a unique value of \(\tau\). Thus \eqref{eq:C-nonzero-condition} removes no physical solution; it merely prevents denominator clearing from introducing solutions at which one of the parameters vanishes. For a frame \(F\), let \begin{equation} \widetilde E_{F,1},\ldots,\widetilde E_{F,9} \label{eq:C-nine-numerators} \end{equation} be the polynomial numerators of its nine block equations. 

We work in the polynomial ring \begin{equation} R:=\mathbb Q[\tau,b,c,e,f] \label{eq:C-ring} \end{equation} and define the frame ideal: 

\begin{equation} I_F:= \left\langle \widetilde\Phi,\widetilde{\Phi^{\ast}}, \widetilde E_{F,1},\ldots,\widetilde E_{F,9}, \tau bcef-1 \right\rangle \subseteq R. \label{eq:C-frame-ideal} \end{equation} 

The common zeros of the generators of \(I_F\) are therefore precisely the non-zero complex solutions of \(\Phi=\Phi^\ast=0\) and the nine block equations associated with \(F\).

We classify the frames as in Section 4. A frame is \emph{generic} if its nine block equations hold automatically for \(X(b,c,e,f)\), without imposing any additional condition on the parameters. Otherwise it is \emph{conditional}, because its block equations impose additional conditions on the parameters. A conditional frame is algebraically empty if: \begin{equation} I_F=R. \label{eq:C-empty-definition} \end{equation} In this case, its generators have no common complex zero and hence no physical solution. A reduced Gröbner basis equal to $\{1\}$ certifies this algebraic emptiness.

For each remaining conditional frame, we determine whether its complex solution set is positive-dimensional or zero-dimensional. A zero-dimensional frame has only finitely many complex solutions. The classification is carried out exactly using reduced Gröbner bases with degree-reverse-lexicographic order and variable order: \begin{equation} \tau>b>c>e>f. \label{eq:C-variable-order} \end{equation}  For each non-empty frame, the exact calculation determines whether its complex solution set is positive-dimensional or zero-dimensional, and provides the additional relations used in the classification below.

\subsection{Exhaustive frame classification}
\label{appC:classification}

Exact reduction of all nine block numerators for all $225$ frames gives three generic frames:
\begin{equation}
 \mathfrak G:=\{(8,8),(10,10),(14,14)\}.
 \label{eq:C-generic-frames}
\end{equation}
These are the frames induced by the simultaneous permutations $p_1,p_2,p_3$ from Section 2.

Set:
\begin{equation}
 A:=\{0,4,12\},
 \qquad
 B:=\{1,5,6\},
 \qquad
 E:=\{2,3,9\}.
 \label{eq:C-Dita-index-sets}
\end{equation}
The 27 Di\c{t}\u{a} frames are:
\begin{equation}
 \mathfrak D:= (A \times A) \sqcup (B \times B) \sqcup (E \times E)
 \label{eq:C-Dita-frames}
\end{equation}
The direct and mixed $C$ frames are:
\begin{equation}
 \mathfrak C_{\mathrm{dir}}
 :=\{(7,7),(11,11),(13,13)\},
 \label{eq:C-direct-frames}
\end{equation}
\begin{equation}
 \mathfrak C_{\mathrm{mix}}
 :=\{(7,11),(7,13),(11,7),(11,13),(13,7),(13,11)\}.
 \label{eq:C-mixed-frames}
\end{equation}
The zero-dimensional $C$ frames split into three groups:
\begin{equation}
\begin{aligned}
 \mathfrak C_{0,1}:={}&
 \{(7,8),(7,10),(7,14),(8,7),(10,7),(14,7)\},\\
 \mathfrak C_{0,2}:={}&
 \{(8,11),(10,11),(11,8),(11,10),(11,14),(14,11)\},\\
 \mathfrak C_{0,3}:={}&
 \{(8,13),(10,13),(13,8),(13,10),(13,14),(14,13)\},
\end{aligned}
\label{eq:C-zero-frame-groups}
\end{equation}
and we put
\begin{equation}
 \mathfrak C_0
 :=\mathfrak C_{0,1}\sqcup\mathfrak C_{0,2}\sqcup\mathfrak C_{0,3}.
 \label{eq:C-zero-frames}
\end{equation}
Finally, define the empty sector by the exact complement
\begin{equation}
 \mathfrak E
 :=\{0,\ldots,14\}^2
 \setminus
 \bigl(
 \mathfrak G\cup\mathfrak D\cup\mathfrak C_{\mathrm{dir}}
 \cup\mathfrak C_{\mathrm{mix}}\cup\mathfrak C_0
 \bigr).
 \label{eq:C-empty-frames}
\end{equation}
This gives a compact, unambiguous specification of every one of the $168$ empty frames.

The corresponding count is
\begin{equation}
 225=3+168+27+3+6+18.
 \label{eq:C-six-sector-count}
\end{equation}

For every $F\in\mathfrak E$, the reduced Gr\"obner basis of $I_F$ contains $1$. So all $168$ empty frames do not have physical solutions.  The six sets in Table~\ref{tab:C-full-frame-classification} are pairwise disjoint and exhaustive.

\begin{longtable}{@{}p{0.25\textwidth}p{0.10\textwidth}p{0.28\textwidth}p{0.27\textwidth}@{}}
\caption{Exact classification of the $225$ frames.}
\label{tab:C-full-frame-classification}\\
\toprule
\textbf{Frame type} & \textbf{Count} & \textbf{Definition} & \textbf{Outcome}\\
\midrule
\endfirsthead
\toprule
\textbf{Frame type} & \textbf{Count} & \textbf{Definition} & \textbf{Outcome}\\
\midrule
\endhead
\bottomrule
\endfoot
Generic & $3$ & $\mathfrak G$ in~\eqref{eq:C-generic-frames} & $C_6^{(1)}$ for repeated-boundary\\
Empty conditional & $168$ & $\mathfrak E$ in~\eqref{eq:C-empty-frames} & No algebraic solution\\
Di\c{t}\u{a} conditional & $27$ & $\mathfrak D=A^2\sqcup B^2\sqcup E^2$ & $D_6^{(1)}$\\
Direct conditional & $3$ & $\mathfrak C_{\mathrm{dir}}$ & $C_6^{(1)}$\\
Mixed conditional & $6$ & $\mathfrak C_{\mathrm{mix}}$ & $C_6^{(1)}$\\
Zero-dim conditional & $18$ & $\mathfrak C_0$ in~\eqref{eq:C-zero-frames} & $C_6^{(1)}$\\
\midrule
Total & $225$ &  & \\
\end{longtable}

  Lemma~4.2 showed that every repeated-boundary presentation supported on a generic frame belongs, up to standard equivalence, to \(C_6^{(1)}\).

\subsection{The $27$ Di\c{t}\u{a} frames}
\label{appC:Dita}
The $27$ frames in $\mathfrak D$ have nine distinct reduced Gröbner bases. Under the parameter transformation \begin{equation} T(b,c,e,f):=(c,b,e^{-1},f^{-1}), \qquad X\bigl(T(b,c,e,f)\bigr)=X(b,c,e,f)^{\mathsf T}, \label{eq:C-transpose-identity} \end{equation} these reduce to six prototype types: \begin{equation} D_{00},\quad D_{04},\quad D_{11},\quad D_{15},\quad D_{22},\quad D_{23}. \label{eq:C-Dita-prototypes} \end{equation} 

Note that these are types of frame ideals, not individual matrix-equivalence classes. Table~\ref{tab:C-Dita-frame-ledger} gives the complete reduction. A direct frame has the same ideal as the named prototype. For a transpose-related frame, applying $T$ to its parameter locus gives the prototype locus. 

\begin{table}[htbp] \centering \small \caption{Reduction of the $27$ Diţă frames to six prototypes.} \label{tab:C-Dita-frame-ledger} \begin{tabular}{@{}lll@{}} \toprule \textbf{Prototype} & \textbf{Direct frames} & \textbf{Transpose-related frames}\\ \midrule $D_{00}$ & $(0,0),(5,5),(9,9)$ & ---\\ $D_{04}$ & $(0,4),(5,6),(9,2)$ & $(4,0),(6,5),(2,9)$\\ $D_{11}$ & $(1,1),(3,3),(12,12)$ & ---\\ $D_{15}$ & $(1,5),(3,9),(12,0)$ & $(5,1),(9,3),(0,12)$\\ $D_{22}$ & $(2,2),(4,4),(6,6)$ & ---\\ $D_{23}$ & $(2,3),(4,12),(6,1)$ & $(3,2),(12,4),(1,6)$\\ \bottomrule \end{tabular} \end{table}

The simplified Laurent ideals and their complete physical parametrisations are given in Table~\ref{tab:C-Dita-prototypes}, where $|z|=1$ and $\varepsilon\in\{1,-1\}$. \begin{longtable}{@{}lll@{}} \caption{Diţă prototype ideals and physical parametrisations.} \label{tab:C-Dita-prototypes}\\ \toprule \textbf{Type} & \textbf{Laurent ideal} & \textbf{$(b,c,e,f)=\Psi_{a,\varepsilon}(z)$}\\ \midrule \endfirsthead \toprule \textbf{Type} & \textbf{Laurent ideal} & \textbf{$(b,c,e,f)=\Psi_{a,\varepsilon}(z)$}\\ \midrule \endhead \bottomrule \endfoot $D_{00}$ & $\langle f-b^2,\ bc+1,\ e^2+b^2\rangle$ & $(z,-z^{-1},\varepsilon iz,z^2)$\\ $D_{04}$ & $\langle e-c^2,\ b+e,\ f^2+e\rangle$ & $(-z^2,z,z^2,\varepsilon iz)$\\ $D_{11}$ & $\langle bc+1,\ e-c^2,\ f^2+e\rangle$ & $(-z^{-1},z,z^2,\varepsilon iz)$\\ $D_{15}$ & $\langle b-f^2,\ ef+1,\ c^2+b\rangle$ & $(z^2,\varepsilon iz,-z^{-1},z)$\\ $D_{22}$ & $\langle bc+1,\ ef+1,\ bf+ce\rangle$ & $(\varepsilon iz^{-1},\varepsilon iz,-z^{-1},z)$\\ $D_{23}$ & $\langle b+c^2,\ bf+1,\ e^2+f\rangle$ & $(-z^2,z,\varepsilon iz^{-1},z^{-2})$\\ \end{longtable} To see that these parametrisations exhaust the physical solutions, choose $z=b$ for $D_{00}$, $z=c$ for $D_{04},D_{11},D_{23}$, and $z=f$ for $D_{15},D_{22}$. The displayed generators then determine the remaining parameters, with precisely the two choices $\varepsilon=\pm1$. Conversely, substitution of each parametrisation gives \begin{equation} \Phi\bigl(\Psi_{a,\varepsilon}(z)\bigr) = \Phi^\ast\bigl(\Psi_{a,\varepsilon}(z)\bigr) =0. \label{eq:C-Dita-Phi} \end{equation} 

Thus every displayed branch is physical for $|z|=1$. For one-based row and column orders $r=(r_1,\ldots,r_6)$ and $s=(s_1,\ldots,s_6)$, write: \begin{equation} X[r,s]_{jk}:=X_{r_j,s_k}. \label{eq:C-Dita-index-orders} \end{equation} Table~\ref{tab:C-Dita-certificates} gives explicit certificates of the form: \begin{equation} D_L(z)\, X\bigl(\Psi_{a,\varepsilon}(z)\bigr)[r,s] = D(q(z)). \label{eq:C-Dita-certificate-form} \end{equation} 
where $[r,s]$ means explicitly reordering the rows and columns, and $D_L$ rephases the reordered rows. For these prototype certificates, the right diagonal factor is the identity. \begingroup \small \setlength{\tabcolsep}{3pt} \begin{longtable}{@{}llllll@{}} \caption{Standard-equivalence certificates for the six prototypes.} \label{tab:C-Dita-certificates}\\ \toprule \textbf{Type} & $\varepsilon$ & $q$ & \textbf{Rows $r$} & \textbf{Columns $s$} & \textbf{Diagonal entries of $D_L$}\\ \midrule \endfirsthead \toprule \textbf{Type} & $\varepsilon$ & $q$ & \textbf{Rows $r$} & \textbf{Columns $s$} & \textbf{Diagonal entries of $D_L$}\\ \midrule \endhead \bottomrule \endfoot $D_{00}$ & $+1$ & $z^{-3}$ & $(1,5,4,6,2,3)$ & $(5,1,3,2,6,4)$ & $(1,-1,i,-i,-i,i)$\\ $D_{00}$ & $-1$ & $iz^{-3}$ & $(1,5,3,2,6,4)$ & $(1,5,3,2,6,4)$ & $(1,1,1,1,1,1)$\\ $D_{04}$ & $+1$ & $z^3$ & $(1,5,4,6,2,3)$ & $(6,3,2,1,4,5)$ & $(1,-i,iz^3,-1,i,-iz^3)$\\ $D_{04}$ & $-1$ & $z^{-3}$ & $(1,5,6,3,4,2)$ & $(6,3,1,5,2,4)$ & $(1,i,-1,iz^3,-iz^3,-i)$\\ $D_{11}$ & $+1$ & $z^{-3}$ & $(1,6,4,5,3,2)$ & $(6,1,2,3,5,4)$ & $(1,-1,i,-i,-i,i)$\\ $D_{11}$ & $-1$ & $z^3$ & $(1,6,5,2,4,3)$ & $(6,1,3,4,2,5)$ & $(1,-1,i,-i,-i,i)$\\ $D_{15}$ & $+1$ & $z^{-3}$ & $(1,6,2,3,5,4)$ & $(3,4,1,2,6,5)$ & $(1,i,-1,iz^3,-iz^3,-i)$\\ $D_{15}$ & $-1$ & $z^3$ & $(1,6,3,4,2,5)$ & $(3,4,2,5,1,6)$ & $(1,-i,-iz^3,i,-1,iz^3)$\\ $D_{22}$ & $+1$ & $z^3$ & $(1,4,3,5,2,6)$ & $(1,4,3,5,2,6)$ & $(1,1,1,1,1,1)$\\ $D_{22}$ & $-1$ & $z^{-3}$ & $(1,4,2,3,6,5)$ & $(1,4,2,3,6,5)$ & $(1,1,1,1,1,1)$\\ $D_{23}$ & $+1$ & $z^3$ & $(1,4,6,5,2,3)$ & $(5,3,2,1,6,4)$ & $(1,-i,iz^3,-1,i,-iz^3)$\\ $D_{23}$ & $-1$ & $z^{-3}$ & $(1,4,5,3,6,2)$ & $(5,3,1,4,2,6)$ & $(1,i,-1,iz^3,-iz^3,-i)$\\ \end{longtable} \endgroup Each certificate is an entrywise identity in $\mathbb Q(i,z)$, and every diagonal entry of $D_L(z)$ is unimodular for $|z|=1$. Hence these identities establish standard Hadamard equivalence on the entire physical branches, including special parameter values. For example, the $D_{22}$ prototype requires only simultaneous row and column permutations: \begin{equation} \begin{aligned} X(iz^{-1},iz,-z^{-1},z)[r_+,r_+] &=D(z^3),\\ X(-iz^{-1},-iz,-z^{-1},z)[r_-,r_-] &=D(z^{-3}), \end{aligned} \label{eq:C-Dita-example} \end{equation} where \begin{equation} r_+=(1,4,3,5,2,6), \qquad r_-=(1,4,2,3,6,5). \end{equation} It remains to cover the transpose-related frames. The displayed Diţă representative satisfies \begin{equation} D(q)^{\mathsf T}=D(q^{-1}). \label{eq:C-Dita-transpose} \end{equation} Let $X=X(b,c,e,f)$ belong to a transpose-related frame. Then $X^{\mathsf T}$ belongs to the corresponding prototype locus, so its certificate has the form \begin{equation} D_L(z)\,X^{\mathsf T}[r,s]=D(q(z)). \end{equation} Transposing this complete identity gives \begin{equation} X[s,r]\,D_L(z)=D(q(z)^{-1}). \label{eq:C-Dita-transposed-certificate} \end{equation} This is again a standard-equivalence certificate: the row and column orders are exchanged, and the left diagonal factor becomes a right diagonal factor. No assumption that transposition itself is standard equivalence is required. The grouped frame ledger and the two branches of each prototype exhaust all physical solutions of the $27$ frames. Therefore every physical point of every frame in $\mathfrak D$ is standard-equivalent to D(q) and belongs to $D_6^{(1)}$.

\subsection{Direct and zero-dimensional $C$ frames}
\label{appC:direct-zero-C}

For the three direct frames, reduction by the corresponding reduced Gröbner bases gives: \begin{equation} g_1\in I_{(7,7)} \qquad g_2\in I_{(11,11)} \qquad g_3\in I_{(13,13)} \label{eq:C-direct-memberships} \end{equation} 

from Appendix B. For the zero-dimensional frames, the exact membership ledger is \begin{equation} g_1\in I_F \qquad \text{for every }F\in\mathfrak C_{0,1}, \label{eq:C-zero-membership-1} \end{equation} \begin{equation} g_2\in I_F \qquad \text{for every }F\in\mathfrak C_{0,2}, \label{eq:C-zero-membership-2} \end{equation} \begin{equation} g_3\in I_F \qquad \text{for every }F\in\mathfrak C_{0,3}. \label{eq:C-zero-membership-3} \end{equation} Each statement in \eqref{eq:C-direct-memberships}--\eqref{eq:C-zero-membership-3} is certified by zero normal form modulo the relevant reduced Gröbner basis. Every physical solution therefore satisfies the corresponding \(g_j=0\). Lemma~4.2 identifies the physical loci \(g_j=0\) with the three chart presentations of \(C_6^{(1)}\). Therefore every physical point of the three direct frames and all eighteen zero-dimensional frames belongs to \(C_6^{(1)}\).

The zero-dimensional frames admit another classification: \begin{equation} \mathfrak C_0 = \mathfrak C_0^{+}\sqcup\mathfrak C_0^{-} \label{eq:C-zero-splitting} \end{equation} 

where: \begin{equation} \begin{aligned} \mathfrak C_0^{+} =\{& (7,8),(7,10),(7,14), (11,8),(11,10),(11,14),\\ & (13,8),(13,10),(13,14) \}, \end{aligned} \label{eq:C-zero-plus-frames} \end{equation} and \begin{equation} \begin{aligned} \mathfrak C_0^{-} =\{& (8,7),(10,7),(14,7), (8,11),(10,11),(14,11),\\ & (8,13),(10,13),(14,13) \}. \end{aligned} \label{eq:C-zero-minus-frames} \end{equation} 

Let $\omega=e^{2\pi i/3} $ then the physical solutions of every frame in \(\mathfrak C_0^{+}\) represent the equivalence class of:  \begin{equation} H_{+}:=C(1,\omega) = \begin{pmatrix} 1&1&1&1&1&1\\ 1&\omega&\omega^2&\omega&1&\omega^2\\ 1&\omega^2&\omega&\omega&\omega^2&1\\ 1&1&1&-1&-1&-1\\ 1&\omega&\omega^2&-\omega&-1&-\omega^2\\ 1&\omega^2&\omega&-\omega&-\omega^2&-1 \end{pmatrix}. \label{eq:C-zero-H-plus} \end{equation}

Similarly, the physical solutions of every frame in \(\mathfrak C_0^{-}\) represent the equivalence class of  \begin{equation} H_{-}:=C(\omega,1)=H_{+}^{\mathsf T} \label{eq:C-zero-minus-equivalence} \end{equation} 

Note that ``zero-dimensional'' refers to the algebraic solution set of the block equations for a particular conditional frame. It does not mean that \(H_{+}\) or \(H_{-}\) is isolated in \(C_6^{(1)}\). 

\subsection{The six mixed $C$ frames}
\label{appC:mixed}

All six mixed frames can be treated uniformly using a phase $\rho$ satisfying $ \rho^3=1$. The frame ideals imply the relations in Table~\ref{tab:C-mixed-sources}, so every physical mixed-frame solution has one of the displayed source forms. The remaining parameters are unit phases satisfying the physical two-circulant constraint.

Each displayed relation has zero normal form modulo the corresponding reduced Gröbner basis. In the first two rows, choose $\rho=e^2$ and $\rho=f^2$, respectively; in the remaining rows, choose $\rho=e,c,f,b$, respectively. Thus $\rho=1$ covers the unit branches, while the other two cube roots cover the primitive-cubic branches, as set out below. \begin{equation} \begin{aligned} \iota&=(1,2,3,4,5,6),\\ p&=(2,3,1,6,4,5),\\ q&=(3,1,2,5,6,4). \end{aligned} \label{eq:C-mixed-orders} \end{equation} 

\begin{table}[htbp] \centering \small \setlength{\tabcolsep}{4pt} \caption{Source and target for all six physical mixed-frame solutions} \label{tab:C-mixed-sources} \begin{tabular}{@{}llll@{}} \toprule \textbf{Frame} & \textbf{Implied relations} & \textbf{Source} & \textbf{Target}\\ \midrule $(7,11)$ & $e^3=1,\ c=e^2$ & $X(b,\rho,\rho^2,f)$ & $g_1=0$\\ $(7,13)$ & $f^3=1,\ b=f^2$ & $X(\rho,c,e,\rho^2)$ & $g_1=0$\\ $(11,7)$ & $e^3=1,\ b=e$ & $X(\rho,c,\rho,f)$ & $g_2=0$\\ $(11,13)$ & $c^3=1,\ e=cf$ & $X(b,\rho,\rho f,f)$ & $g_2=0$\\ $(13,7)$ & $f^3=1,\ c=f$ & $X(b,\rho,e,\rho)$ & $g_3=0$\\ $(13,11)$ & $b^3=1,\ f=be$ & $X(\rho,c,e,\rho e)$ & $g_2=0$\\ \bottomrule \end{tabular} \end{table}

The six sources satisfy the following standard-equivalence identities: \begin{equation} \begin{aligned} L_{7,11}\, X(b,\rho,\rho^2,f)[\iota,p] &= X\left(\frac{\rho}{b},\frac1b, \frac1f,\frac{\rho^2}{f}\right),\\ L_{7,13}\, X(\rho,c,e,\rho^2)[\iota,q] &= X\left(\frac1c,\frac{\rho}{c}, \frac{\rho^2}{e},\frac1e\right),\\ L_{11,7}\, X(\rho,c,\rho,f)[\iota,q] &= X\left(\frac{\rho}{c},\frac1c,f,\rho\right),\\ L_{11,13}\, X(b,\rho,\rho f,f)[\iota,p] &= X\left(\frac{\rho}{b},\frac1b,\frac1f,\rho\right),\\ L_{13,7}\, X(b,\rho,e,\rho)[\iota,p] &= X\left(\frac{\rho}{b},\frac1b,\rho^2,\rho^2e\right),\\ X(\rho,c,e,\rho e)[p,\iota]\,R_{13,11} &= X\left(\frac1c,\frac{\rho}{c}, \frac{\rho^2}{e},\rho^2\right). \end{aligned} \label{eq:C-mixed-certificates} \end{equation} Here \begin{equation} \begin{aligned} L_{7,11} &=\operatorname{diag} (1,b\rho,b^2\rho^2,bf,b\rho,b\rho^2/f),\\ L_{7,13} &=\operatorname{diag} (1,c^2\rho^2,c\rho,ce,c\rho^2/e,c\rho),\\ L_{11,7} &=\operatorname{diag} (1,c^2\rho^2,c\rho,c\rho,c\rho^2f,c/f),\\ L_{11,13} &=\operatorname{diag} (1,b\rho,b^2\rho^2,bf,b/(\rho f),b\rho),\\ L_{13,7} &=\operatorname{diag} (1,b\rho,b^2\rho^2,b\rho,b/e,be/\rho),\\ R_{13,11} &=\operatorname{diag} (1,c\rho,c^2\rho^2,c/(\rho e),ce,c\rho). \end{aligned} \label{eq:C-mixed-diagonals} \end{equation} All six identities hold entrywise using only $\rho^3=1$. Their denominators are monomials in non-zero parameters, and their diagonal factors are unitary on the physical locus. Thus they give standard equivalences at every physical source point. In particular, the last identity uses the displayed row order and a right diagonal factor, without treating transposition as an equivalence operation. Each target has unit-modulus parameters and satisfies $\Phi=0$ whenever its source does. Direct substitution also gives the target equations listed in Table~\ref{tab:C-mixed-sources}. For example, the first target satisfies \begin{equation} \frac{\rho}{b}\frac{\rho^2}{f} = \frac1{bf} = \frac1b\frac1f, \end{equation} so $g_1=0$. Lemma~4.2 therefore identifies every target with a member of $C_6^{(1)}$. Since the source forms cover every physical solution of all six mixed frames, every such solution represents a member of $C_6^{(1)}$.

\subsection{Conclusion and proof of Lemma~4.3}
\label{appC:conclusion}

The conditional frames form the disjoint union \begin{equation} \{0,\ldots,14\}^2\setminus\mathfrak G = \mathfrak E\sqcup\mathfrak D \sqcup\mathfrak C_{\mathrm{dir}} \sqcup\mathfrak C_{\mathrm{mix}} \sqcup\mathfrak C_0. \end{equation} The frames in $\mathfrak E$ are algebraically empty. Section~\ref{appC:Dita} places every physical solution of a frame in $\mathfrak D$ in $D_6^{(1)}$, while Sections~\ref{appC:direct-zero-C} and~\ref{appC:mixed} place every physical solution of the remaining conditional frames in $C_6^{(1)}$. Thus every physical solution supported on a conditional frame belongs, up to standard equivalence, to $D_6^{(1)}\cup C_6^{(1)}$. This applies in particular to the transported frame of any repeated-boundary Karlsson presentation, proving Lemma~4.3. \qed

\end{document}